\documentclass[prb,twocolumn,showpacs]{revtex4}
\usepackage{graphicx} 

\usepackage{amsmath,amsthm,amssymb,mathrsfs}
\usepackage{bm}

\begin{document}

\title{Thermally activated switching rate of a nanomagnet in the presence of spin torque}

\author{Tomohiro Taniguchi$^{1}$, Yasuhiro Utsumi$^{2}$, and Hiroshi Imamura$^{1}$}
 \affiliation{
 $^{1}$ National Institute of Advanced Industrial Science and Technology (AIST), Spintronics Research Center, Tsukuba, Ibaraki 305-8568, Japan, \\
 $^{2}$ Faculty of Engineering, Mie University, Tsu, Mie, 514-8507, Japan.
 }

 \date{\today} 
 \begin{abstract}
  {
    The current dependence of the spin torque switching rate in a thermally activated region
    of an in-plane magnetized system was studied. 
    The logarithm of the switching rate depended nonlinearly on current in the high-current region, 
    $I_{\rm c} \lesssim I < I_{\rm c}^{*}$, 
    where $I_{\rm c}$ and $I_{\rm c}^{*}$ are critical currents 
    distinguishing the stability of the magnetization. 
    We also found that the attempt frequency had a minimum around $I_{\rm c}$, 
    and that the attempt frequency at $I_{\rm c}$ was three orders of magnitude 
    smaller than that at zero current, 
    contrary to the assumption in previous analyses of experiments that it remains constant. 
  }
 \end{abstract}

 \maketitle


\section{Introduction}
\label{sec:Introduction}

The escape problem of a Brownian particle 
from a meta-stable state is 
ubiquitous in many fields of science, 
such as chemical reaction of molecules 
\cite{hanggi86,hanggi90,gardiner09,bell78,dekker86,hummer03,husson08}. 
Spin-torque-driven magnetization switching \cite{slonczewski96,berger96} 
in nanostructured ferromagnets in the thermally activated region also belongs to this problem, 
which has been extensively studied 
because of its potential application to spintronics devices 
such as magnetic random access memory (MRAM) and magnetic sensor. 
The observation of the magnetization switching 
provides us important information, 
such as the retention time of the MRAM. 
More than a decade has passed since the first experimental and theoretical works 
on spin torque switching in the thermally activated region 
\cite{albert02,yakata09,bedau10,cheng10,koch04,li04,apalkov05}. 


The spin torque switching can be regarded as 
the Brownian motion in the presence of a non-conservative force, 
contrary to the switching by magnetic field, 
which is a conservative force defined as the gradient of a potential. 
The lack of a general method 
to formulate the switching rate in the presence of the non-conservative force 
is an unresolved problem in statistical physics \cite{stein89,maier92}. 
Therefore, many assumptions have been used 
in the previous theories of the spin torque switching \cite{koch04,li04,apalkov05}. 
However, recent works 
\cite{suzuki09,pinna12,pinna13,taniguchi11a,taniguchi12b,taniguchi13} have revealed 
the limits of the applicability of 
previous theories. 
For example, 
the switching rate has been assumed to obey the Arrhenius law, $\nu=f e^{-\Delta}$, 
with linear scaling of the switching barrier, 
$\Delta=\Delta_{0}(1-I/I_{\rm c})$, 
where $f$, $\Delta_{0}$, $I$, and $I_{\rm c}$ are 
the attempt frequency, the thermal stability, the current, 
and the critical current of the precession around the easy axis, respectively \cite{koch04,li04,apalkov05}. 
However, the linear scaling is valid only in the low-current region \cite{taniguchi13}, 
while a relatively large current has been applied 
in experiments \cite{albert02,yakata09,bedau10} 
to observe the switching quickly. 
The use of the linear scaling leads to an error 
of the estimation of the thermal stability \cite{taniguchi11a}. 
Another issue is that the transition state theory previously adopted \cite{apalkov05} 
cannot estimate the switching rate 
under a low damping limit \cite{coffey12}, 
while the Gilbert damping constant 
of materials typically used in spintronics application 
are very low \cite{oogane06}, 
i.e., $\alpha=10^{-3}-10^{-2}$. 
These facts prompted us to revisit 
the theory of spin torque switching 
in a thermally activated region. 


In this paper, 
we study the spin-torque-switching rate 
of an in-plane magnetized system 
using the mean first passage time approach 
\cite{coffey12,hanggi90,gardiner09,taniguchi12}. 
The introduction of the effective energy density 
enables us to calculate the switching rate 
even in the presence of the non-conservative force. 
The switching rate showed a nonlinear dependence on the current 
in the high-current region ($I_{\rm c} \lesssim I < I_{\rm c}^{*}$) 
on a logarithmic scale, 
where $I_{\rm c}^{*} \simeq 1.27 I_{\rm c}$ is the spin torque switching current 
at zero temperature. 
The attempt frequency was strongly suppressed around $I_{\rm c}$ 
contrary to the assumption in previous experimental analysis that 
it remains constant \cite{albert02,yakata09,bedau10}. 
For example, the attempt frequency at $I_{\rm c}$ was 
three orders of magnitude smaller than 
that at the zero current. 
The theoretical approach presented in this paper is useful 
for the escape problem of a Brownian particle 
under a non-conservative force 
when the magnitude of the non-conservative force is much smaller than 
that of the conservative force.


The paper is organized as follows. 
In Sec. \ref{sec:Fokker-Planck Equation in Energy Space}, 
the Fokker-Planck equation for the magnetization dynamics 
in the energy space is introduced 
based on the small damping assumption. 
In Sec. \ref{sec:Mean First Passage Time Approach to Switching Rate}, 
the current dependence of the switching rate, 
as well as that of the attempt frequency, is calculated 
by using the mean first passage time approach. 
Section \ref{sec:Conclusion} is devoted to the conclusion.



\begin{figure}
\centerline{\includegraphics[width=1.0\columnwidth]{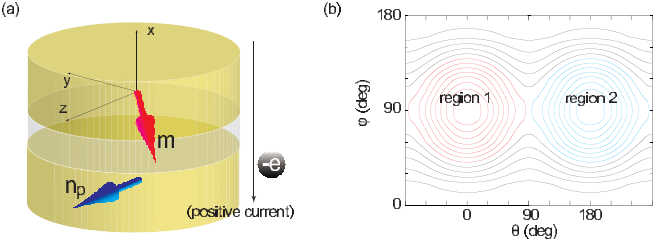}}\vspace{-3.0ex}
\caption{
         (a) A schematic view of the in-plane magnetized system. 
             The unit vectors pointing in the magnetization directions of the free and the pinned layers are 
             denoted as $\mathbf{m}$ and $\mathbf{n}_{\rm p}$, respectively. 
             The $x$-axis is normal to the film plane 
             and the $z$-axis is parallel to the in-plane easy axis. 
         (b) A schematic view of the constant energy curves. 
             Two low-energy regions are separated by a saddle point. 
             The area outside of regions 1 and 2 (black lines) 
             corresponds to the high-energy region. 
         \vspace{-3ex}}
\label{fig:fig1}
\end{figure}


\section{Fokker-Planck Equation in Energy Space}
\label{sec:Fokker-Planck Equation in Energy Space}

Figure \ref{fig:fig1} (a) schematically shows 
an in-plane magnetized system, 
where the $x$ and $z$ axes are 
normal to the film plane 
and parallel to the in-plane easy axis, respectively. 
The unit vectors pointing in the magnetization directions 
of the free and the pinned layers are denoted as 
$\mathbf{m}=(\sin\theta\cos\varphi,\sin\theta\sin\varphi,\cos\theta)$ 
and $\mathbf{n}_{\rm p}=\mathbf{e}_{z}$, respectively. 
Here, we assume that the magnetization dynamics is well described by the macrospin model. 
The macrospin assumption is, at least for the grand state, guaranteed by 
the spin torque diode experiment \cite{tulapurkar05} 
in which the oscillation frequency of the free layer magnetization agrees with 
the ferromagnetic resonance frequency derived by the macrospin model. 
The positive current is defined as the electron flow 
from the free layer to the pinned layer. 
The energy density of the in-plane magnetized system is 
\begin{equation}
\begin{split}
  E
  &=
  -\frac{MH_{\rm K}}{2}
  \left(
    \mathbf{m}
    \cdot
    \mathbf{e}_{z}
  \right)^{2}
  +
  \frac{4 \pi M^{2}}{2}
  \left(
    \mathbf{m}
    \cdot
    \mathbf{e}_{x}
  \right)^{2}, 
  \label{eq:energy}
\end{split}
\end{equation}
where $M$, $H_{\rm K}$, and $-4\pi M$ are 
the magnetization, 
the uniaxial anisotropy field along the $z$-axis, 
and the demagnetization field along the $x$-axis, 
respectively. 
The magnetic field is defined as $\mathbf{H}=-\partial E/(\partial M \mathbf{m})$. 
Figure \ref{fig:fig1} (b) schematically shows 
the constant energy curves of $E$ in $(\theta,\varphi)$ space. 
The in-plane magnetized system has two low-energy regions 
around the energy minima at $\mathbf{m}=\pm\mathbf{e}_{z}$ 
corresponding to $E=-MH_{\rm K}/2 \equiv E_{\rm K}$. 
These two low-energy regions are separated by the saddle point 
$\mathbf{m}=\pm\mathbf{e}_{y}$ 
at which the energy density $E_{\rm s}=0$. 
We named the low-energy region, $E_{\rm K} \le E \le E_{\rm s}$, 
around $\mathbf{m}=+\mathbf{e}_{z}(-\mathbf{e}_{z})$ as region 1 (2). 
The area outside regions 1 and 2 corresponds to 
the high-energy region. 
The magnetization dynamics is described by the Landau-Lifshitz-Gilbert equation 
with the random torque, 
\begin{equation}
\begin{split}
  \frac{d \mathbf{m}}{dt}
  =&
  -\gamma
  \mathbf{m}
  \times
  \mathbf{H}
  -
  \gamma
  H_{\rm s}
  \mathbf{m}
  \times
  \left(
    \mathbf{n}_{\rm p}
    \times
    \mathbf{m}
  \right)
\\
  &-
  \gamma
  \mathbf{m}
  \times
  \mathbf{h}
  +
  \alpha
  \mathbf{m}
  \times
  \frac{d \mathbf{m}}{dt},
  \label{eq:LLG}
\end{split}
\end{equation}
where $\gamma$ and $\alpha$ are 
the gyromagnetic ratio and the Gilbert damping constant, respectively. 
The spin torque strength, $H_{\rm s}=\hbar \eta I/(2eMV)$, 
consists of the current $I$, the spin polarization $\eta$, 
and the volume of the free layer. 
The components of the random field, $h_{k}$ ($k=x,y,z$), 
satisfy the fluctuation-dissipation theorem \cite{brown63},
$\langle h_{i}(t) h_{j}(t^{\prime}) \rangle = (2D/\gamma^{2})\delta_{ij}\delta(t-t^{\prime})$, 
where the diffusion coefficient $D=\alpha \gamma k_{\rm B}T/(MV)$ 
consists of the Boltzmann constant $k_{\rm B}$, 
and the temperature $T$. 


During the switching between the regions 1 and 2, 
the magnetization precesses on the constant energy curve 
around the easy axis. 
The precession period is determined by the anisotropy fields, $H_{\rm K}$ and $4\pi M$, 
and is typically on the order of 1 ns \cite{tulapurkar05}. 
On the other hand, 
the switching time is determined by the damping, the spin torque, and the random field, 
and is on the order of 1 $\mu$s-1 ms, depending on the current magnitude \cite{yakata09}. 
Such long time scale of the switching is due to the fact that 
the correlation function of the random torque, 
which induces the switching, 
is proportional to the small parameter $\alpha$ \cite{oogane06}. 
Therefore, the precession period on the constant energy curve is 
much shorter than the switching time. 
Also, because of the large demagnetization field due to the thin film geometry, 
as soon as the energy exceeds the saddle points energy, 
the magnetization moves from region 1 (2) to 2 (1) 
by the precession around the demagnetization field, 
and relaxes to region 2 (1).
Therefore, the dominant contribution to the switching rate 
is the time climbing the potential well of 
region 1 or 2. 
Thus, we average the magnetization dynamics on the constant energy curve 
in regions 1 and 2, 
and neglect the high-energy region. 
The averaged dynamics is described by 
the Fokker-Planck equation in the energy space \cite{dykman12}, 
which can be derived from Eq. (\ref{eq:LLG}) 
and is given by 
\begin{equation}
  \frac{\partial \mathcal{P}}{\partial t}
  +
  \frac{\partial J}{\partial E}
  =
  0,
  \label{eq:Fokker-Planck-energy-P}
\end{equation}
where $\mathcal{P}=\mathcal{P}(E,t|E^{\prime},t^{\prime})$ is the transition probability function 
of the magnetization direction from the state $(E^{\prime},t^{\prime})$ to $(E,t)$. 
The probability current is 
\begin{equation}
\begin{split}
  J
  =&
  -\frac{\alpha M \mathscr{M}_{\alpha}}{\gamma \tau}
  \frac{d \mathscr{E}}{dE}
  \mathcal{P}
  -
  D 
  \left(
    \frac{M}{\gamma}
  \right)^{2}
  \mathscr{M}_{\alpha}
  \frac{\partial}{\partial E}
  \frac{\mathcal{P}}{\tau}. 
  \label{eq:current-energy-P}
\end{split}
\end{equation}
We use the approximation $1+\alpha^{2} \simeq 1$  
because the present theory is based on 
the small damping assumption. 
The effective energy density for the region $i$ is defined as 
\begin{equation}
  \mathscr{E}_{i}(E)
  =
  \int_{E_{\rm s}}^{E} 
  d E^{\prime}
  \left[
    1
    -
    \frac{\mathscr{M}_{\rm s}(E^{\prime})}{\alpha \mathscr{M}_{\alpha}(E^{\prime})}
  \right],
  \label{eq:effective_energy_1}
\end{equation}
where the lower boundary of the integral, $E_{\rm s}$, is chosen 
to make the effective energy density continuous 
at the boundary of the regions 1 and 2. 
The precession period on the constant energy curve, $\tau=\oint dt$, 
and the functions 
$\mathscr{M}_{\alpha}=\gamma^{2}\oint dt [\mathbf{H}^{2}-(\mathbf{m}\cdot\mathbf{H})^{2}]$ 
and $\mathscr{M}_{\rm s}=\gamma^{2} H_{\rm s} \oint dt [\mathbf{n}_{\rm p}\cdot\mathbf{H}-(\mathbf{m}\cdot\mathbf{n}_{\rm p})(\mathbf{m}\cdot\mathbf{H})]$, 
which are proportional to 
the energy dissipation due to the damping 
and the work done by spin torque on the constant energy curve, respectively, are given by 
\begin{equation}
  \tau
  =
  \frac{4}{\gamma \sqrt{H_{\rm K}(4\pi M-2E/M)}}
  \mathsf{K}
  \left(
    \sqrt{
      \frac{4\pi M(H_{\rm K}+2E/M)}{H_{\rm K}(4\pi M-2E/M)}
    }
  \right),
  \label{eq:period}
\end{equation}
\begin{equation}
\begin{split}
  \mathscr{M}_{\alpha}
  &=
  4 \gamma
  \sqrt{
    \frac{4\pi M-2E/M}{H_{\rm K}}
  }
  \left[
    \frac{2E}{M}
    \mathsf{K}
    \left(
      \sqrt{
        \frac{4\pi M(H_{\rm K}+2E/M)}{H_{\rm K}(4\pi M-2E/M)}
      }
    \right)
  \right.
\\
  &\ \  +
  \left.
    H_{\rm K}
    \mathsf{E}
    \left(
      \sqrt{
        \frac{4\pi M(H_{\rm K}+2E/M)}{H_{\rm K}(4\pi M-2E/M)}
      }
    \right)
  \right],
  \label{eq:Melnikov_alpha}
\end{split}
\end{equation}
\begin{equation}
\begin{split}
  \mathscr{M}_{\rm s}
  &=
  \pm
  \frac{2\pi \gamma H_{\rm s} (H_{\rm K}+2E/M)}{\sqrt{H_{\rm K}(H_{\rm K}+4\pi M)}}, 
  \label{eq:Melnikov_s}
\end{split}
\end{equation}
where $\mathsf{K}(k)$ and $\mathsf{E}(k)$ are 
the first and second kind of 
complete elliptic integrals, respectively. 
The double sign in Eq. (\ref{eq:Melnikov_s}) 
means the upper ($+$) for region 1 and the lower ($-$) for region 2. 
This difference of the sign of $\mathscr{M}_{\rm s}$ 
represents the fact that 
the spin torque for $I>0$ destabilizes the magnetization in region 1 
while it stabilizes the magnetization in region 2. 


\begin{figure}
\centerline{\includegraphics[width=1.0\columnwidth]{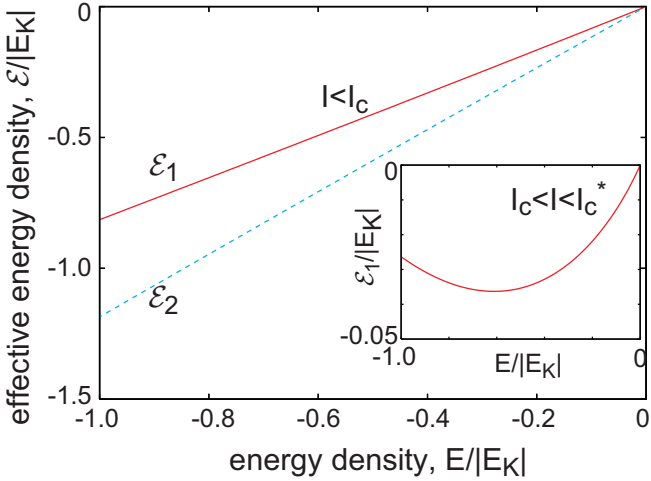}}\vspace{-3.0ex}
\caption{
         Schematic views of the effective energy density of region 1 
         ($E_{\rm K} \le E \le E_{\rm s}$) 
         in the energy space 
         for $I = 0.2 I_{\rm c}$. 
         Both $E$ and $\mathscr{E}_{1}$ are normalized by $|E_{\rm K}|=MH_{\rm K}/2$. 
         The inset shows $\mathscr{E}_{1}$ for $I=0.2 (I_{\rm c}^{*}-I_{\rm c})+I_{\rm c}$. 
         The dotted line is $\mathscr{E}_{2}$. 
         \vspace{-3ex}}
\label{fig:fig2}
\end{figure}


Equation (\ref{eq:current-energy-P}) indicates that, 
after averaging the magnetization dynamics on the constant energy curve, 
the switching can be regarded as the Brownian motion on the effective energy density 
in which the equation of motion with the deterministic force is 
$(1/\tau)\oint dt (dE/dt)=-[\alpha M \mathscr{M}_{\alpha}/(\gamma \tau)](d \mathscr{E}_{i}/d E)$. 
The thermally activated region is defined by $-d \mathscr{E}/dE < 0$. 
We note that 
\begin{equation}
  \lim_{E \to E_{\rm K}} 
  \frac{d \mathscr{E}_{i}}{dE}
  =
  1 
  \mp 
  \frac{I}{I_{\rm c}},
  \label{eq:E_Ic}
\end{equation}
\begin{equation} 
  \lim_{E \to E_{\rm s}} 
  \frac{d \mathscr{E}_{i}}{dE} 
  = 
  1 
  \mp 
  \frac{I}{I_{\rm c}^{*}}, 
  \label{eq:E_Ic*}
\end{equation}
respectively, 
where 
$I_{\rm c}=[2\alpha eMV/(\hbar\eta)](H_{\rm K}+2\pi M)$ 
and $I_{\rm c}^{*}=[4 \alpha eMV/(\pi \hbar \eta)]\sqrt{4\pi M (H_{\rm K}+4\pi M)} \simeq 1.27 I_{\rm c}$ \cite{taniguchi13,hillebrands06}. 
The physical meanings of $I_{\rm c}$ and $I_{\rm c}^{*}$ are that, 
for region 1, 
the state $\mathbf{m}=\mathbf{e}_{z}$ is destabilized by the current $I>I_{\rm c}$ 
while the magnetization switches without the thermal fluctuation for $I>I_{\rm c}^{*}$. 
Therefore, in terms of the current, 
the thermally activated region is defined by $I<I_{\rm c}^{*}$. 
It should also be noted that the steady state solution of Eq. (\ref{eq:Fokker-Planck-energy-P}) in the region $i$ is 
the Boltzmann distribution with the effective energy density, 
i.e., $\mathcal{P}/\tau \propto e^{-\mathscr{E}_{i}(E)V/(k_{\rm B}T)}$. 


Figure \ref{fig:fig2} shows 
the typical dependences of $\mathscr{E}_{1}$ on $E$ 
for $I \le I_{\rm c}$ and $I_{\rm c} < I \le I_{\rm c}^{*}$, 
where the values of the parameters \cite{taniguchi13} are 
$M=1000$ emu/c.c., 
$H_{\rm K}=200$ Oe, 
$V=\pi \times 80 \times 35 \times 2.5$ nm${}^{3}$, 
$\eta=0.8$, 
$\gamma=1.764 \times 10^{7}$ rad/(Oe$\cdot$s), 
and $\alpha=0.01$, respectively. 
We denote the energy density corresponding to 
the local minimum of the effective energy density as $E^{*}$, 
which for region 1 is located at 
\begin{equation}
  E^{*}({\rm region\ 1})
  =
  \begin{cases}
    E_{\rm K} & (I \le I_{\rm c}) \\
    {\rm solution\ of\ }d \mathscr{E}_{1}/dE=0 & (I_{\rm c}  < I < I_{\rm c}^{*}). 
  \end{cases}
  \label{eq:E*}
\end{equation}
The minimum of $\mathscr{E}_{2}$ 
always locates at $E^{*}=E_{\rm K}$. 


\section{Mean First Passage Time Approach to Switching Rate}
\label{sec:Mean First Passage Time Approach to Switching Rate}

The mean first passage time \cite{hanggi90,gardiner09,taniguchi12}, 
which characterizes how long the magnetization 
stays in the energy range $E^{*} \le E \le E_{\rm s}$ of the region $i$, 
is defined as 
\begin{equation}
  \mathcal{T}_{i}(E)
  =
  \int_{0}^{\infty} 
  dt 
  \int_{E^{*}}^{E_{\rm s}} 
  d E_{1} 
  \mathcal{P}(E_{1},t|E,0). 
\end{equation}
The equation to determine the mean first passage time is obtained 
from the adjoint of Eq. (\ref{eq:Fokker-Planck-energy-P}), 
and is given by 
\begin{equation}
\begin{split}
  \frac{\alpha M \mathscr{M}_{\alpha}}{\gamma \tau}
  \frac{d \mathscr{E}_{i}}{dE}
  \frac{d \mathcal{T}_{i}}{dE}
  -
  D 
  \left(
    \frac{M}{\gamma}
  \right)^{2}
  \frac{1}{\tau}
  \frac{d}{dE}
  \mathscr{M}_{\alpha}
  \frac{d \mathcal{T}_{i}}{dE}
  =
  1.
  \label{eq:MFPT_equation}
\end{split}
\end{equation}
We use the reflecting and the absorbing boundary conditions \cite{hanggi90,gardiner09,taniguchi12} 
at $E=E^{*}$ and $E=E_{\rm s}$, respectively: 
that is, $d \mathcal{T}_{i}(E^{*})/d E=0$ and $\mathcal{T}_{i}(E_{\rm s})=0$. 
Then, the mean first passage time is given by 
\begin{equation}
\begin{split}
  \mathcal{T}_{i}(E)
  =
  \frac{\gamma V}{\alpha M k_{\rm B}T}
  &
  \int_{E}^{E_{\rm s}} 
  d E_{1} 
  \int_{E^{*}}^{E_{1}} 
  d E_{2} 
  \frac{\tau(E_{2})}{\mathscr{M}_{\alpha}(E_{1})}
\\
  & \times
  \exp
  \left\{
    \frac{[\mathscr{E}_{i}(E_{1})-\mathscr{E}_{i}(E_{2})]V}{k_{\rm B}T}
  \right\}.
  \label{eq:MFPT}
\end{split}
\end{equation}
The switching rate from region $i$ to region $j$ is 
\begin{equation}
  \nu_{ij}
  =
  \frac{ d \mathscr{E}_{j}(E_{\rm s})/d E}{ d \mathscr{E}_{i}(E_{\rm s}) / d E + d \mathscr{E}_{j}(E_{\rm s}) / d E}
  \frac{1}{\mathcal{T}_{i}(E^{*})}. 
  \label{eq:switching_rate}
\end{equation}
Here, we assume that 
once the magnetization reaches the saddle point, 
the probability of it moving to the regions $i$ or $j$ is proportional to 
the gradient of the effective energy, 
i.e., the deterministic force acting on a Brownian particle \cite{comment3}.
For a conservative system \cite{hanggi90}, 
Eq. (\ref{eq:switching_rate}) is $1/(2 \mathcal{T}_{i})$. 
The switching probability $P$ 
and the switching current distribution $dP/dI$ 
measured in the experiments 
can be calculated from Eq. (\ref{eq:switching_rate}). 
For example, for $I>0$, 
the switching probability from $\mathbf{m}=\mathbf{e}_{z}$ to $\mathbf{m}=-\mathbf{e}_{z}$ is 
$P \simeq 1-e^{-\int_{0}^{t} \nu_{12}(t^{\prime}) dt^{\prime}}$. 
It should be noted that $\lim_{I \to I_{\rm c}^{*}} \mathcal{T}_{1}(E^{*})=0$ 
because region 1 is no longer stable 
due to the spin torque; 
thus, the magnetization immediately switches to region 2. 
For the same reason, 
$\lim_{I \to I_{\rm c}^{*}} \nu_{21} = 0$. 


\begin{figure}
\centerline{\includegraphics[width=0.8\columnwidth]{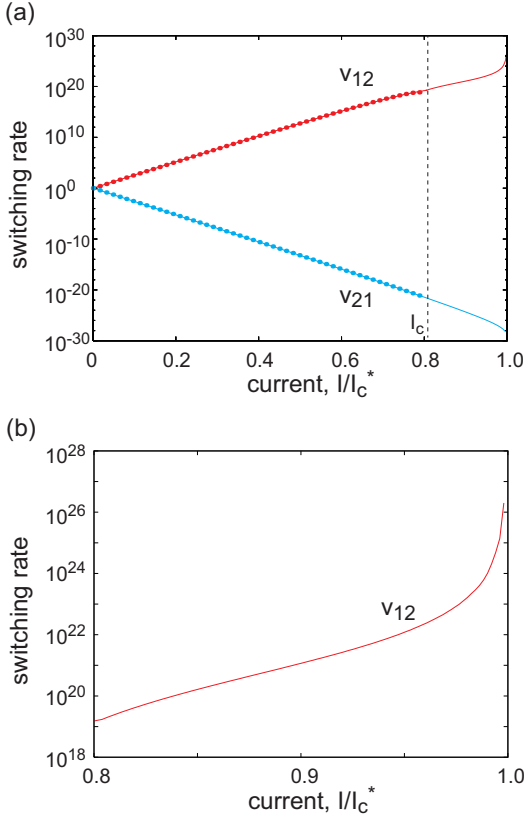}}\vspace{-3.0ex}
\caption{
         (a) Dependence of the switching rate $\nu_{ij}$ 
             on the current $I$. 
             The values are normalized by those at $I=0$, 
             while the current is normalized by $I_{\rm c}^{*}$. 
             The dots represent the analytical solutions 
             derived in the region $I<I_{\rm c}$. 
             The dashed line represents the position of $I_{\rm c}$ ($I_{\rm c}/I_{\rm c}^{*}\simeq 0.81$). 
         (b) An enlarged view of the switching rate in the high-current region, $I_{\rm c} < I < I_{\rm c}^{*}$. 
         \vspace{-3ex}}
\label{fig:fig3}
\end{figure}



Equations (\ref{eq:MFPT}) and (\ref{eq:switching_rate}) indicate that 
the switching rate cannot be expressed as the Arrhenius law, in general. 
However, it is convenient to introduce the switching barrier and the attempt frequency as 
\begin{equation}
  \Delta_{i}
  =
  \frac{[\mathscr{E}_{i}(E_{\rm s})-\mathscr{E}_{i}(E^{*})]V}{k_{\rm B}T},
  \label{eq:switching_barrier}
\end{equation}
\begin{equation}
  f_{ij}
  =
  \nu_{ij}
  e^{\Delta_{i}}.
  \label{eq:attempt_frequency}
\end{equation}
The current dependence of the switching barrier was extensively 
studied in Ref. \cite{taniguchi13}. 
In the low-current region, $I < I_{\rm c}$, 
corresponding to the high barrier limit, $\Delta_{i} \gg 1$ \cite{comment1}, 
the exponential terms 
in Eq. (\ref{eq:MFPT}) are 
dominated by $E_{1}=E_{\rm s}$ and $E_{2}=E_{\rm K}$, respectively. 
Using the Taylor expansion of $\mathscr{E}_{i}$, 
$\mathcal{T}_{i}$ can be approximated as 
\begin{equation}
\begin{split}
  \mathcal{T}_{i}(E_{\rm K})
  &
  \simeq
  \frac{\gamma k_{\rm B}T \tau(E_{\rm K}) e^{\Delta_{i}}}{\alpha MV \mathscr{M}_{\alpha}(E_{\rm s}) [d\mathscr{E}_{i}(E_{\rm K})/dE][d\mathscr{E}_{i}(E_{\rm s})/dE]},
  \label{eq:MFPT_approx}
\end{split}
\end{equation}
where 
$\mathscr{M}_{\alpha}(E_{\rm s})=4\gamma \sqrt{H_{\rm K}4\pi M}$ 
and $\tau(E_{\rm K})=2\pi/[\gamma\sqrt{H_{\rm K}(H_{\rm K}+4\pi M)}]$. 
Then, the switching rate obeys the Arrhenius law as follows: 
\begin{equation}
\begin{split}
  \nu_{ij}
  =&
  \frac{\alpha MV \mathscr{M}_{\alpha}(E_{\rm s})}{2\gamma k_{\rm B}T \tau(E_{\rm K})}
  \left(
    1
    \mp
    \frac{I}{I_{\rm c}}
  \right)
  \left[
    1
    -
    \left(
      \frac{I}{I_{\rm c}^{*}}
    \right)^{2}
  \right]
\\
  & \times
  \exp
  \left[
    -\Delta_{0}
    \left(
      1
      \mp
      \frac{I}{\tilde{I}_{\rm c}}
    \right)
  \right],
  \label{eq:rate_Brown}
\end{split}
\end{equation}
where $\Delta_{0}=MH_{\rm K}V/(2k_{\rm B}T)$ is the thermal stability. 
The term $(1\mp I/I_{\rm c})$ of Eq. (\ref{eq:rate_Brown}) 
arises from $d \mathscr{E}_{1}(E_{\rm K})/d E$ in Eq. (\ref{eq:MFPT_approx}) 
while the term $[1 - (I/I_{\rm c}^{*})^{2}]$ of Eq. (\ref{eq:rate_Brown}) 
arises from $d \mathscr{E}_{i}(E_{\rm s})/dE$ in 
Eqs. (\ref{eq:switching_rate}) and (\ref{eq:MFPT_approx}), 
respectively. 
The current $\tilde{I}_{\rm c}$ is defined as 
$I/\tilde{I}_{\rm c}=\int_{E_{\rm K}}^{E_{\rm s}} (dE/|E_{\rm K}|) \mathscr{M}_{\rm s}/(\alpha \mathscr{M}_{\alpha})$, 
which satisfies $I_{\rm c}<\tilde{I}_{\rm c}<I_{\rm c}^{*}$. 
Although the linear scaling of the switching barrier appears 
in this low-current region, 
the scaling current is not the switching current, 
as argued in Refs. \cite{koch04,li04,apalkov05}. 
This means that the previous analyses of the experiments 
underestimate the real value of the switching current \cite{albert02,yakata09,koch04,li04,taniguchi13a}. 
Equation (\ref{eq:rate_Brown}) can be directly reproduced 
by applying Brown's approach \cite{brown63} 
to Eq. (\ref{eq:Fokker-Planck-energy-P}), 
as shown in Appendix. 


Equation (\ref{eq:rate_Brown}) becomes zero in the zero-dissipation limit ($\alpha \to 0$) 
because the correlation function of 
the thermal field, which induces the switching, 
is proportional to $\alpha$ 
according to the fluctuation-dissipation theorem. 
However, the switching rate 
based on the transition state theory \cite{apalkov05}
given by $\nu_{ij}=\exp[-\Delta_{0}(1 \mp I/I_{\rm c})]/\tau(E_{\rm K})$ 
remains finite in the zero-dissipation limit. 
This problem has already been pointed out in the case of the magnetic field switching \cite{coffey12}. 
The terms except for $1/\tau(E_{\rm K})$ in Eq. (\ref{eq:rate_Brown}) can be 
regarded as correction terms to the transition state theory 
used in Ref. \cite{apalkov05}. 


Figure \ref{fig:fig3} (a) shows 
the dependence of the switching rate $\nu_{ij}$ 
on the current 
numerically obtained from Eq. (\ref{eq:MFPT}), 
where $\nu_{ij}$ are normalized by the values at $I=0$. 
The values of the parameters are those used in Fig. \ref{fig:fig2} 
with $T=300$ K. 
The analytical solution, Eq. (\ref{eq:rate_Brown}), for $I<I_{\rm c}$ is shown by dots, 
and shows good agreement with the numerical result. 
The switching rate in the high-current region, 
$I_{\rm c}<I<I_{\rm c}^{*}$, is shown in Fig. \ref{fig:fig3} (b). 
One of the main results in this paper is 
the nonlinear dependence of $\nu_{ij}$ in the relatively high-current region 
on the logarithmic scale, 
while the linear dependence has been widely used 
in previous works \cite{koch04,li04}
by assuming the linear scaling of the switching barrier 
and the constant attempt frequency. 


\begin{figure}
\centerline{\includegraphics[width=1.0\columnwidth]{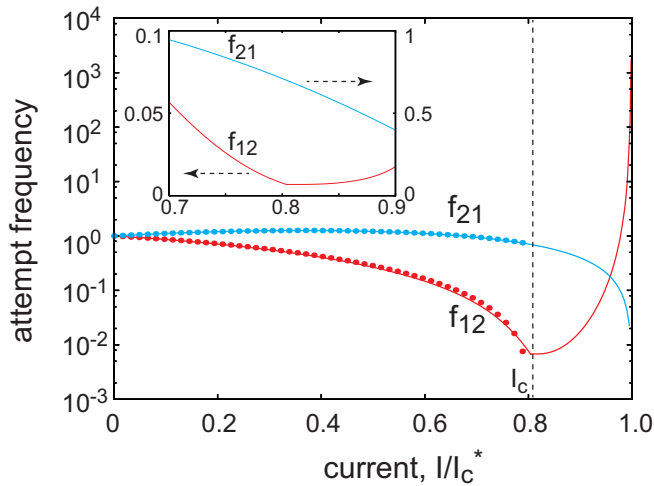}}\vspace{-3.0ex}
\caption{
         Dependence of the attempt frequency, $f_{ij}$, on the current.  
         The values are normalized by those at $I=0$, 
         while the current is normalized by $I_{\rm c}^{*}$. 
         The dots represent the analytical solutions obtained from Eq. (\ref{eq:rate_Brown}). 
         The dashed line represents the position of $I_{\rm c}$ ($I_{\rm c}/I_{\rm c}^{*}\simeq 0.81$). 
         The inset shows the linear plots of $f_{12}$ and $f_{21}$ 
         around $I_{\rm c}$. 
         \vspace{-3ex}}
\label{fig:fig4}
\end{figure}



The current dependence of the attempt frequency, $f_{ij}$, is shown 
in Fig. \ref{fig:fig4}. 
The attempt frequency, $f_{12}$, decreases with increasing current 
for $I \lesssim I_{\rm c}$, 
while it increases for $I_{\rm c} \lesssim I < I_{\rm c}^{*}$. 
The discontinuity of the slope of $f_{12}$ around $I_{\rm c}$ arises 
for the following reason. 
According to Eqs. (\ref{eq:switching_rate}) and (\ref{eq:MFPT_approx}), 
the attempt frequency for $I<I_{\rm c}$ is approximately proportional to 
the gradient of $\mathscr{E}_{1}$ at its minimum, $d \mathscr{E}_{1}(E_{\rm K})/dE$, 
which decreases with increasing current. 
Here, $d \mathscr{E}_{1}(E_{\rm K})/dE$ arises from 
the Taylor expansion of $\mathscr{E}_{1}$ in Eq. (\ref{eq:MFPT}). 
On the other hand, 
$d \mathscr{E}_{1}/dE$ for $ I_{\rm c} \lesssim I < I_{\rm c}^{*}$ 
is zero at the minimum of $\mathscr{E}_{1}$ 
as shown by Eq. (\ref{eq:E*}). 
Then, $f_{12}$ is approximately proportional to 
the curvature of $\mathscr{E}_{1}$ at its minimum, $d^{2}\mathscr{E}_{1}(E^{*})/dE^{2}$, 
which increases with increasing the current. 
Thus, the attempt frequency shows a minimum around $I_{\rm c}$. 
The attempt frequency in Fig. \ref{fig:fig4} strongly depends on the current. 
For example, 
the attempt frequency at $I_{\rm c}$ is 
three orders of magnitude smaller than 
that at zero current. 
Contrary to this result, 
the attempt frequency has been generally assumed to be constant 
in previous experimental analyses \cite{albert02,yakata09,bedau10,cheng10,koch04,li04}. 


\section{Conclusion}
\label{sec:Conclusion}

In summary, 
the spin-torque-switching rate of an in-plane magnetized system was studied. 
The current dependence of the switching rate was obtained numerically, 
and an analytical formula in the low-current region was derived. 
The logarithm of the switching rate depends
nonlinearly on the current in the high-current region. 
The switching barrier linearly depends on the current 
in the low-current region, 
which guarantees the validity of the previous theories \cite{koch04,li04,apalkov05}, 
although the scaling current 
$\tilde{I}_{\rm c}$ 
is not identical to the switching current. 
The attempt frequency has a minimum around the critical current $I_{\rm c}$,
and exhibits a strong current dependence 
while it has been assumed to be constant in previous experimental analyses. 


\section*{Acknowledgement}

The authors would like to acknowledge 
M. Marthaler, D. S. Golubev, H. Sukegawa, T. Yorozu, 
H. Kubota, H. Maehara, A. Emura, T. Nozaki, M. Konoto, K. Yakushiji, A. Fukushima, K. Ando, and S. Yuasa 
for the valuable discussions they had with us. 
This work was supported by JSPS KAKENHI Grant-in-Aid for Young Scientists (B) 25790044. 


\appendix
\section{Brown's Approach to Eq. (\ref{eq:rate_Brown})}
\label{sec:Appendix_A}

The switching rate in the high barrier limit, Eq. (\ref{eq:rate_Brown}), can be obtained 
by using the Brown's approach \cite{brown63}. 
To this end, 
it is convenient to use $W= M \mathcal{P}/(\gamma \tau)$, instead of $\mathcal{P}$. 
In terms of $W$, 
Eqs. (\ref{eq:Fokker-Planck-energy-P}) and (\ref{eq:current-energy-P}) can be expressed as 
\begin{equation}
  \frac{\gamma \tau}{M}
  \frac{\partial W}{\partial t}
  +
  \frac{\partial J}{\partial E}
  =
  0,
\end{equation}
\begin{equation}
  J
  =
  -\frac{\alpha k_{\rm B}T}{V}
  \mathscr{M}_{\alpha}
  e^{-\mathscr{E}V/(k_{\rm B}T)}
  \frac{\partial}{\partial E}
  e^{\mathscr{E}V/(k_{\rm B}T)}
  W.
  \label{eq:current}
\end{equation}
To guarantee $\Delta_{i} \gg 1$, 
we assume that $I < I_{\rm c}$, 
i.e., $E^{*}=E_{\rm K}$. 
Then, $\Delta_{i}$ is given by $\Delta=\Delta_{0}(1 \mp I / \tilde{I}_{\rm c})$. 
In the high barrier limit, 
the probability functions near the minima of $\mathscr{E}$ are 
approximately the Boltzmann distribution functions, 
while a tiny constant flow of the probability current 
crosses over the saddle point. 
The probability functions of the region 1 and 2 
around $E_{\rm K}$ are expressed as 
\begin{equation}
  W_{i}(E)
  =
  W_{i}(E_{\rm K})
  \exp
  \left\{
    -\frac{[\mathscr{E}_{i}(E)-\mathscr{E}_{i}(E_{\rm K})]V}{k_{\rm B}T}
  \right\},
\end{equation}
where $i=1,2$, 
$W_{i}(E_{\rm K})=W_{i}(E_{\rm s}) \exp\{[\mathscr{E}_{i}(E_{\rm s})-\mathscr{E}_{i}(E_{\rm K})]V/(k_{\rm B}T)\}$, 
and $W_{i}(E_{\rm s})$ is the probability function 
at the saddle point 
satisfying $W_{1}(E_{\rm s})=W_{2}(E_{\rm s})$. 
Since the probability functions show sharp peaks 
around $E_{\rm K}$, 
and rapidly decreases by approaching to $E_{\rm s}$, 
the integrals of the probability functions, 
\begin{equation}
  n_{i}
  =
  \frac{\gamma}{M}
  \int_{E_{\rm K}}^{E_{i}} 
  d E 
  W_{i}(E) 
  \tau(E), 
  \label{eq:n-def}
\end{equation}
are independent of the upper boundaries, 
$E_{i}$, 
which are arbitrary points 
located in the region 1 and 2 close to $E_{\rm s}$. 
Equation (\ref{eq:n-def}) can be expressed as 
$n_{i}=W_{i}(E_{\rm K}) \exp [\mathscr{E}_{i}(E_{\rm K})V/(k_{\rm B}T)]I_{i}$, 
where $I_{i}$ are given by 
\begin{equation}
  I_{i}
  =
  \frac{\gamma}{M}
  \int_{E_{\rm K}}^{E_{1}} 
  d E 
  \exp
  \left[
    -\frac{\mathscr{E}_{i}(E)V}{k_{\rm B}T}
  \right]
  \tau(E). 
  \label{eq:integral_I}
\end{equation}
The exponential term in Eq. (\ref{eq:integral_I}) rapidly decreases 
from $E=E_{\rm K}$ to $E=E_{\rm s}$. 
Thus, by using the Taylor expansion of $\mathscr{E}_{i}$ around $E=E_{\rm K}$ 
and using the fact that 
$\lambda_{E}^{*}=-d\mathscr{E}_{i}/dE \neq 0$ 
for $I<I_{\rm c}$,
$I_{i}$ can be approximated to 
\begin{equation}
\begin{split}
  I_{i}
  &\simeq
  \frac{\gamma k_{\rm B}T \tau(E_{\rm K})}{MV [d \mathscr{E}_{i}(E_{\rm K})/dE]}
  e^{-\mathscr{E}_{i}(E_{\rm K})V/(k_{\rm B}T)}.
  \label{eq:integral_I_approx}
\end{split}
\end{equation}
The double sign means 
the upper for region 1 and the lower for region 2.


Next, we consider the flow of the probability current 
from region 1 to region 2 
crossing the saddle point. 
Equation (\ref{eq:current}) can be rewritten as 
\begin{equation}
  \frac{JV}{\alpha k_{\rm B}T \mathscr{M}_{\alpha}}
  e^{\mathscr{E}(E)V/(k_{\rm B}T)}
  =
  -\frac{\partial}{\partial E}
  e^{\mathscr{E}(E)V/(k_{\rm B}T)}
  W.
  \label{eq:current-energy-different-form}
\end{equation}
By assuming the divergenceless current \cite{brown63}, 
the integral of Eq. (\ref{eq:current-energy-different-form}) 
over $[E_{1},E_{\rm s}]$ is given by 
\begin{equation}
\begin{split}
  &
  \frac{JV}{\alpha k_{\rm B}T}
  \int_{E_{1}}^{E_{\rm s}} 
  d E 
  \frac{e^{\mathscr{E}_{1}(E)V/(k_{\rm B}T)}}{\mathscr{M}_{\alpha}}
\\
  &=
  W_{1}(E_{1})
  e^{\mathscr{E}_{1}(E_{1})V/(k_{\rm B}T)}
  -
  W_{1}(E_{\rm s})
  e^{\mathscr{E}_{1}(E_{\rm s})V/(k_{\rm B}T)}.
\end{split}
\end{equation}
We also integrate Eq. (\ref{eq:current-energy-different-form}) 
over $[E_{2},E_{\rm s}]$ 
by changing the sign of the probability current $J$. 
Then, we obtain the following equation; 
\begin{equation}
  \frac{JV}{\alpha k_{\rm B}T}
  I_{\alpha}
  =
  W_{1}(E_{1})
  e^{\mathscr{E}_{1}(E_{1})V/(k_{\rm B}T)}
  -
  W_{2}(E_{2})
  e^{\mathscr{E}_{2}(E_{2})V/(k_{\rm B}T)},
  \label{eq:I_alpha_equation}
\end{equation}
where the right hand side is identical to $(n_{1}/I_{1})-(n_{2}/I_{2})$. 
On the other hand, $I_{\alpha}$ is given by 
\begin{equation}
\begin{split}
  I_{\alpha}
  &=
  \int_{E_{1}}^{E_{\rm s}} 
  d E 
  \frac{e^{\mathscr{E}_{1}(E)V/(k_{\rm B}T)}}{\mathscr{M}_{\alpha}}
  +
  \int_{E_{2}}^{E_{\rm s}} 
  d E 
  \frac{e^{\mathscr{E}_{2}(E)V/(k_{\rm B}T)}}{\mathscr{M}_{\alpha}}
\\
  &\simeq 
  \frac{k_{\rm B}T}{\mathscr{M}_{\alpha}(E_{\rm s})V}
  \left[
    \frac{e^{\mathscr{E}_{1}(E_{\rm s})V/(k_{\rm B}T)}}{d \mathscr{E}_{1}(E_{\rm s})/dE}
    +
    \frac{e^{\mathscr{E}_{2}(E_{\rm s})V/(k_{\rm B}T)}}{d \mathscr{E}_{2}(E_{\rm s})/dE}
  \right].
  \label{eq:I_alpha}
\end{split}
\end{equation}


The probability current satisfies 
$d n_{1}/dt = -dn_{2}/dt = -J$. 
Thus, we obtain the following rate equation 
between the region 1 and 2; 
\begin{equation}
  \frac{d n_{1}}{dt}
  =
  -\frac{d n_{2}}{dt}
  =
  -n_{1}
  \nu_{12}
  +
  n_{2}
  \nu_{21},
  \label{eq:rate_equation}
\end{equation}
where the switching rate from the region $i$ to the region $j$ is 
$\nu_{ij}=\alpha k_{\rm B}T/(I_{i}I_{\alpha}V)$. 
By using Eqs. (\ref{eq:E_Ic}), (\ref{eq:E_Ic*}), (\ref{eq:integral_I_approx}), and (\ref{eq:I_alpha}), 
the explicit form of the switching rate is given by 
\begin{equation}
  \nu_{ij}
  =
  \frac{\alpha MV \mathscr{M}_{\alpha}(E_{\rm s})}{2\gamma k_{\rm B}T \tau(E_{\rm K})}
  \left(
    1
    \mp
    \frac{I}{I_{\rm c}}
  \right)
  \left[
    1
    -
    \left(
      \frac{I}{I_{\rm c}^{*}}
    \right)^{2}
  \right]
  e^{-\Delta_{i}},
  \label{eq:switching_rate_Brown}
\end{equation}
where the double sign means 
the upper for $(i,j)=(1,2)$ and the lower for $(2,1)$. 
Equation (\ref{eq:switching_rate_Brown}) is identical to Eq. (\ref{eq:rate_Brown}). 
The solutions of Eq. (\ref{eq:rate_equation}) for constant current in time 
with the initial condition $n_{1}(t=0)=1$ and $n_{2}(t=0)=0$ are given by 
\begin{equation}
  n_{1}
  =
  \frac{\nu_{21}}{\nu_{12}+\nu_{21}}
  +
  \frac{\nu_{12}}{\nu_{12}+\nu_{21}}
  e^{-(\nu_{12}+\nu_{21})t},
\end{equation}
\begin{equation}
  n_{2}
  =
  \frac{\nu_{12}}{\nu_{12}+\nu_{21}}
    -
  \frac{\nu_{12}}{\nu_{12}+\nu_{21}}
  e^{-(\nu_{12}+\nu_{21})t}.
\end{equation}
For the positive current, 
$\nu_{12} \gg \nu_{21}$, 
and therefore $n_{1} \simeq e^{-\nu_{12}t}$ and $n_{2}=1-e^{-\nu_{12}t}$, 
where $n_{2}$ corresponds to the switching probability measured in the experiments. 



\begin{thebibliography}{34}
\bibitem[{com({\natexlab{a}})}]{comment3}
\bibinfo{note}{By using the boundary condition $\partial
  (\mathcal{P}/\tau)/\partial E=0$ and the continuity of $J$ at $E=E_{\rm s}$,
  the probability in the region $i$ close to the saddle point is given by $(d
  \mathscr{E}_{j}/dE)/[(d \mathscr{E}_{i}/dE)+(d \mathscr{E}_{j}/dE)]$, which
  determines the rate moving to the region $j$.}
\bibitem[{com({\natexlab{b}})}]{comment1}
\bibinfo{note}{For MRAM application, the thermal stability $\Delta_{0}$ larger
  than 40 is required. In such sample, the switching barrier at $I=I_{\rm c}$
  is on the order of unity.}
\end{thebibliography}


\end{document}